\documentclass{aa}
\usepackage{afterpage}
\usepackage{graphics}

\def\bc{\begin{center}}
\def\ec{\end{center}}
\def\be{\begin{equation}}
\def\ee{\end{equation}}
\def\bea{\begin{eqnarray}}
\def\eea{\end{eqnarray}}
\def\ve{\varepsilon}
\def\mkm{\mu{\rm m}}
\def\ve{\varepsilon}
\def\bet{{\beta}~\rm Pictoris}
\begin{document}
\title{Modelling the optical properties of composite and porous interstellar grains}
\titlerunning{Composite and porous interstellar grains}

\author{ N.V.~Voshchinnikov\inst{1,2},
         V.B.~Il'in\inst{1,2},
         and
         Th.~Henning\inst{3}
         }
\authorrunning{Voshchinnikov, Il'in, Henning}

\institute{
Sobolev Astronomical Institute,
St.~Petersburg University, Universitetskii prosp. 28,
           St.~Petersburg, 198504 Russia
\and
 Isaac Newton Institute of Chile, St.~Petersburg Branch
\and
Max-Planck-Institut f\"ur Astronomie, K\"onigstuhl 17, D-69117 Heidelberg, Germany
}
\offprints{N.V. Voshchinnikov, nvv@astro. \protect\linebreak
  spbu.ru}

\date{Received $<$date$>$; accepted $<$date$>$}

\abstract{
There are indications that interstellar and interplanetary
dust grains have an  inhomogeneous and fluffy structure. We investigate
different methods to describe light scattering by such composite particles.
Both a model of layered particles and discrete dipole calculations
for particles with Rayleigh and non-Rayleigh inclusions are used.

The calculations demonstrate that porosity is a key parameter for determining
light scattering. We find that the optical properties of the layered particles depend
on the number and position of layers if the number of layers is
small ($\la 15$). For a larger number of layers the scattering characteristics become
independent of the layer sequence. The optical properties of particles 
with inclusions depend on the size of inclusions provided the porosity is large.
The scattering characteristics of  very porous particles with
inclusions of different sizes are found to be close to those of multi-layered spheres.

We compare the results of these calculations with
the predictions of the effective medium theories (EMT)
which are often used in astronomy as a tool to calculate the optical properties of
composite particles. The results of our analysis show that the internal structure of grains
(layers versus inclusions) only slightly affects the optics of particles provided the porosity 
does not exceed 50\%. It is also demonstrated that in this case the optical properties of composite grains 
calculated with EMT agree with the results of the exact method for layered particles.
For larger porosity, the standard EMT rules (i.e., Garnett and Bruggeman rules)
give  reliable results for particles with Rayleigh inclusions only.

\keywords{{\bf\rm Scattering -- dust, extinction -- comets -- interplanetary medium}}
 }     

\maketitle

\section{Introduction}
Various processes operating in interstellar and circumstellar
media are believed to produce inhomogeneous and porous cosmic dust grains
(Dorschner \cite{dor99}, Draine  \cite{d03}).
However, the real structure of interstellar grains remains to be established.
The particles may have a layered structure because of the
their formation in circumstellar environments (Dominik et al. \cite{dsg93})
and further evolution in molecular clouds (Ehrenfreund \cite{ere99},
Greenberg \cite{g99}).
On the other hand, collisions of grains
tend to induce coagulation and partial destruction of particles.
Because of this, interstellar grains should have inclusions of different
size.  Therefore,
considering the optical properties of cosmic dust grains
we are forced to solve the difficult problem of the interaction of
radiation with composite particles of different structure.

Fortunately, scientists
had felt the necessity to treat the scattering by
composite and inhomogeneous particles {\bf\rm or} media
{\bf\rm consisting} of several components
even earlier than
the existence of interstellar dust was established.
 Garnett~(\cite{gar04}) was {\bf\rm the} first to find
the averaged (effective) dielectric functions
of {\bf\rm such} a medium assuming that one material
was a matrix (host material) in which another material was embedded
in {the form of small} inclusions
(so called Maxwell--Garnett mixing rule of the Effective Medium Theory; EMT).
 {\bf\rm Later}, Bruggeman~(\cite{brug35}) deduced another rule which
was symmetric with respect to the materials.
 These classical mixing rules are still the most popular ones.

Many scientific and applied problems require
calculations of light scattering by inhomogeneous particles
with  good accuracy.

 This  first became possible at the beginning of the 1950s when
the Mie solution for homogeneous spheres was
generalized to core-mantle spherical particles in three independent
papers (Aden \& Kerker \cite{ak51},
Shifrin \cite{s52}, G\"uttler \cite{gut52}).
 G\"uttler's solution was used by
Wickramasinghe~(\cite{w63}) who first calculated the extinction of
layered (graphite core-ice mantle) analogues of cosmic grains.

Mathis \&  Whiffen~(\cite{mw89})
introduced the first consistent model of composite cosmic grains which were
very porous (the volume fraction of vacuum $\sim 80$\,\%)
aggregates of small amorphous carbon, silicate and
iron subparticles. The optical properties of such particles
were calculated with the Mie theory and EMT.
Mathis~(\cite{m96}) updated the composite grain model
taking into account the abundances of heavy elements
obtained for cluster and field B stars and young F, G stars
(Snow \& Witt \cite{sw96}).
The new model consisted of three dust grain populations where the visual/near-IR
extinction was explained by aggregates with $\sim 45$\,\% vacuum
in volume.

Now light scattering computations for inhomogeneous (composite) particles with
layers or inclusions from different materials  or aggregate particles
are often made using the discrete dipole approximation (DDA)
or simpler theories
(see Voshchinnikov \cite{v02} for discussion).
{\bf\rm Note that the DDA is a  method which still is computationally demanding.
Therefore, it is mostly used for
 illustrative calculations and not for  mass production (e.g.,
Wolff  et al. \cite{wcms-l94}, \cite{wcg98}, Vaidya  et al. \cite{vgdc01},
Andersen et al. \cite{andetal03}).

The idea to represent composite interstellar grains by multi-layered spheres
(Voshchinnikov \&  Mathis \cite{vm99},
see also Iat\`\i \,  et al. \cite{iatietal01}) has no immediate physical
justification (although such particles may form in stellar envelopes
and molecular clouds),
but is very attractive as an exact method to calculate the optical
properties of composite particles.
Such a model permits us to include an arbitrary
fraction of any material, and computations do not require large
resources. However, the distribution of material inside particles is
always spherically symmetric even when its volume fraction tends to zero.

 In this paper, we compare the optical properties of composite interstellar
grains of various porosity
obtained from calculations for layered spheres,
pseudospherical particles with inclusions and homogeneous spheres with
an effective refractive index.
 The description of the particle models is given in Sect.~\ref{model}.
We {\bf\rm compute} different efficiency factors, albedo, etc.
and analyze how these quantities depend on the order and number of layers
and the size of inclusions  (Sect.~\ref{uni}).
 Special attention is paid to the  consideration
of very porous grains (Sect.~\ref{vac}) because
of {\bf\rm their} particular importance in astronomy,
{\bf\rm for example,}
for {\bf\rm modelling of} comets (Greenberg \& Hage~\cite{grha91})
and the disc of $\bet$ (Li \& Greenberg~\cite{li:gre98}).
The possibility to describe the light scattering by
porous particles using Mie theory
with different EMT rules is studied in Sect.~\ref{emt}.
Concluding remarks are presented in Sect.~\ref{concl}.

\section{Models of composite grains}\label{model}

Processes operating in the winds of late-type stars such as grain nucleation
and growth, shattering in the diffuse interstellar medium, and finally coagulation
and accretion in molecular clouds and protoplanetary disks lead certainly to
dust grains with rather irregular shapes and very complicated internal structure
(Dorschner \& Henning \cite{dh95}). The details of the grain interiors are not directly important for surface chemistry,
but the optical behaviour of the particles may be a strong function of this structure.
Direct evidence for the structure of these particles is difficult to obtain with the
exception of interplanetary dust grains collected in the upper atmosphere and the solar system.
Therefore, a more general attempt to describe the particles and to explore changes
in their optical properties is required.

A frequently used approach in astronomy is the modelling of  inhomogeneous grains
by two-layered (core-mantle) spheres and
particles with voids or inclusions using  EMT-Mie calculations. In this paper,
we consider layered particles and particles with inclusions
as the models  for the description of the optical properties of the
inhomogeneous or composite grains.
The amount of a material in such particles
is determined by its volume  fraction
$V_i$ ($\Sigma_i V_i /V_{\rm total} = 1)$.
The particle porosity ${\cal P}$ ($0 \leq {\cal P} < 1$)
is introduced as
\be
{\cal P} = V_{\rm vac} /V_{\rm total}
= 1 - V_{\rm solid} /V_{\rm total}, \label{por}
\ee
where $V_{\rm vac}$ and $V_{\rm solid}$ is the volume fraction
of vacuum and solid material, respectively.

In our calculations presented below, composite particles of
several materials are considered. The refractive indices for them
were taken from the Jena--Petersburg Database of Optical Constants (JPDOC)
which was described by Henning et al.~(\cite{heal99}) and
J\"ager et al.~(\cite{jetal02}).

 Carbon and silicates are the materials most often used in cosmic dust
models (see Mathis et al. \cite{mrn}, Draine \& Lee \cite{dl84} and so on).
 We consider the particles composed of
amorphous carbon (AC1), astronomical silicate (astrosil) and vacuum
with varied volume fraction of each constituent.
 The optical constants for AC1 ($m=1.98+0.23i$) and astrosil ($m=1.68+0.03i$),
corresponding to the wavelength $\lambda=0.55\,\mkm$,
were taken from the papers of Rouleau \& Martin~(\cite{rm91}) and
Laor \&  Draine~(\cite{laordr93}), respectively.

\subsection{Layered particles}

Here we represent  
composite grains by particles consisting of many concentric
spherical  homogeneous layers of cyclically changing materials.
Such a model is not primarily meant as a physical description
of the actual grain structure, but as a possibility for 
describing light scattering by inhomogeneous particles of complex structure.

 Vacuum can be one of the materials, and a
composite particle may have a central cavity or voids in the form
of concentric layers.
 This model allows one to include at any position inside
a spherical particle  any fraction of a material
(from extremely small to very large).
The light scattering calculations are based on the exact theory
which is true for particles of any size and refractive index.

\begin{figure}[htb]
\bc
\resizebox{\hsize}{!}{\includegraphics{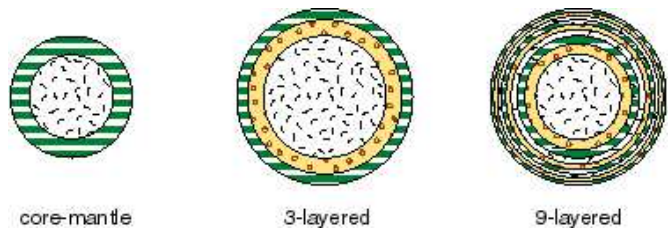}}
\caption{
The cut of the spherical particles by the plane.
The models of composite {\bf\rm grains} containing the same
amount {\bf\rm of} carbon and silicate are shown.
{\bf\rm The 3- and 9-layered spheres consist of equivolume spherical layers
with the total volume fractions of carbon, silicate
and vacuum equal to 33.33\%.}
The core-mantle particle {\bf\rm includes} the same mass of carbon and
silicate but is free of vacuum.
}\label{f00}\ec\end{figure}

The schematic representation
(cross  cut)
of layered spherical  grains is given in Fig.~\ref{f00}.
 {\bf\rm The particles} are composed of a specified number of
concentric spherical homogeneous layers.
 The order of the layers and their total number can be specified
separately.
 Following Voshchinnikov \&  Mathis~(\cite{vm99}),
we assume further that
different material layers cyclically change inside a particle and
call the repeating set of such layers a shell.
 This means that a {\bf\rm particle} consists of a specified number of concentric shells, {\bf\rm and}
the simplest {\bf\rm model}
particle contains one shell of two materials.
 The core-mantle particle {\bf\rm presented in Fig.~\ref{f00} }
does not contain vacuum, but its mass
is the same as that of the other two particles {\bf\rm shown}.
{\bf\rm As a result, its volume is less by factor of 1/3 and the outer radius
by $\sqrt[3]{1/3} \approx 0.69$}, respectively.

The formal solution to the light scattering problem
for $n$-layered spheres can be easily written in  matrix form
with the separation of variables method (see, e.g., Kerker \cite{k69}).
In this case the scattering coefficients are calculated as the ratios
of two determinants of order $2n+1$ containing
Riccati--Bessel functions and their first derivatives of real and complex
arguments.
 However, for practical reasons, it is better to use
the recursive algorithm developed by Wu \& Wang~(\cite{ww91}) and
Johnson~(\cite{j96}). In order to make  calculations for highly
absorbing particles of
{\bf\rm large sizes}, one should {\bf\rm take into account the} modifications
suggested by Wu et al.~(\cite{wuetal97}) and
Gurwich et al.~(\cite{gsk01}).

\subsection{Particles with inclusions}\label{incl}

The optical properties of particles with inclusions can be estimated
on the basis of rather complicated calculations (see, e.g.,
Wolff  et al. \cite{wcms-l94}, Videen \& Ch\'ylek \cite{vidch98})
or from laboratory measurements (Kolokolova \& Gustafson \cite{kolgust01}).
If the volume  fraction of inclusions is not very large ($\la 10$\%),
the EMT-Mie calculations give the results with  good accuracy
(Wolff  et al. \cite{wcms-l94}, \cite{wcg98},
Kolokolova \& Gustafson \cite{kolgust01}).

However, our goal is the consideration of  particles with
an arbitrary amount of inclusions and
different porosity. Therefore,
the calculations  are performed
with the discrete dipole approximation (DDA).
 We use the last version of the DDA program {\bf\rm (DDSCAT 6.0)}
developed by Draine \& Flatau~(\cite{df03}).
 This technique can treat
particles of arbitrary shapes and/or of inhomogeneous structure.
 A detailed review of the DDA and its applications is given by
Draine~(\cite{d00}).

 The particles (``targets'' in the DDSCAT terminology)
are constructed using two special routines.
 One routine produces spherical targets with inclusions of
a fixed size, while another {\bf\rm creates} targets with
a given distribution of inclusions over their sizes.
 Both routines produce first a cube with randomly
distributed cubic inclusions.
 The {\bf\rm sizes} of the target $d_{\rm max}$ and of the inclusions $d_{\rm incl}$
are expressed in units of {\bf\rm the interdipole distance} $d$.
 In the cube, we just inscribe a sphere and remove all inclusions
{\bf\rm and} their parts being {\bf\rm outside} the sphere.


 In contrast to  previous modelling (Henning \& Stognienko \cite{hs93},
 Lumme \& Rahola \cite{lr94},
Wolff  et al. \cite{wcms-l94}, \cite{wcg98}, Vaidya  et al. \cite{vgdc01}),
porous particles are not produced by removing dipoles
or inclusions from a target but {\bf\rm by attributing the refractive index
$m=1.000001+0.0i$ to them}. We believe {\bf\rm that } such a structure better
corresponds to
cosmic aggregates.

For the purpose of treating very porous particles, the number of dipoles
in pseudospheres is taken quite large.
In all cases considered, the particles with
the maximum size $d_{\rm max}=91$ are studied.
 This value corresponds to the total number of dipoles in pseudospheres
$N_{\rm dip}=357128 - 381915$ depending on the
size of inclusions $d_{\rm incl}$.
 Thus, the criterion of the validity of the DDA for extinction/scattering
cross sections $|m|kd < 1$\footnote{Note that sometimes the fulfillment
of this criterion does not guarantee the correct result (see, e.g.,
discussion in Andersen et al. \cite{andetal03}).}
of Draine \& Flatau~(\cite{df03}) ($k=2 \pi/\lambda$
is the wavenumber with $\lambda$ being the wavelength in vacuum)
is satisfied up to the size parameter
$x_{\rm porous}= 2 \pi r /\lambda \approx 27 - 28$.

\begin{figure}[htb]
\bc
\resizebox{\hsize}{!}{\includegraphics{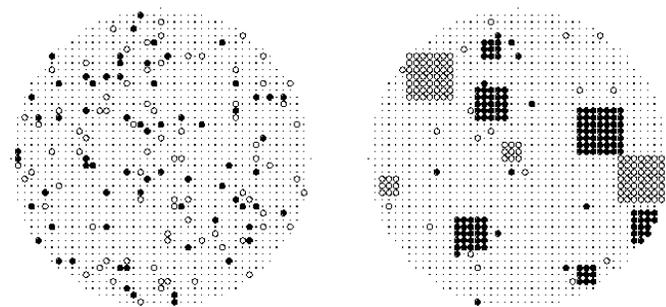}}
\caption{
The cut of the pseudospherical particles with
the maximum size $d_{\rm max}=45$ by the plane.
The models of very porous particles with small single size (left)
and different size (right) cubic inclusions are shown.
The volume fractions of carbon and silicate are equal to 5\%.}
\label{f01}\ec\end{figure}

Targets with the values
of $d_{\rm incl}$ {\bf\rm ranging} from 1 to 9 are considered.
 The resulting structure
(cross cut)
 of pseudospherical grains is shown in Fig.~\ref{f01}.
 Note that the inclusions of the size $d_{\rm incl}=1$ are dipoles, while the inclusions
with $d_{\rm incl}=3, 5, 7$ and 9 consist of 27, 125, 343 and 729 dipoles,
respectively.

 The optical characteristics of pseudospheres {\bf\rm with inclusions} were averaged over three
targets obtained for different random number sets. The calculations
show that in our case such an approach is practically equivalent to
time-consuming numerical averaging over target orientations.

\section{Towards  unified optical properties}\label{uni}

It is evident that  composite particles of various structure and
shape should exist in space.
Here we focus our attention on spheres of different
porosity because of their frequent use  in the
modelling of interstellar and cometary grains and
possible importance
in attacking the  problem of cosmic abundances.
Our analysis should also help to estimate the range of validity
of some previous models and to clarify the  physical background of them.

\subsection{Particles of  moderate porosity}\label{mpor}

\begin{figure}\bc
\resizebox{\hsize}{!}{\includegraphics{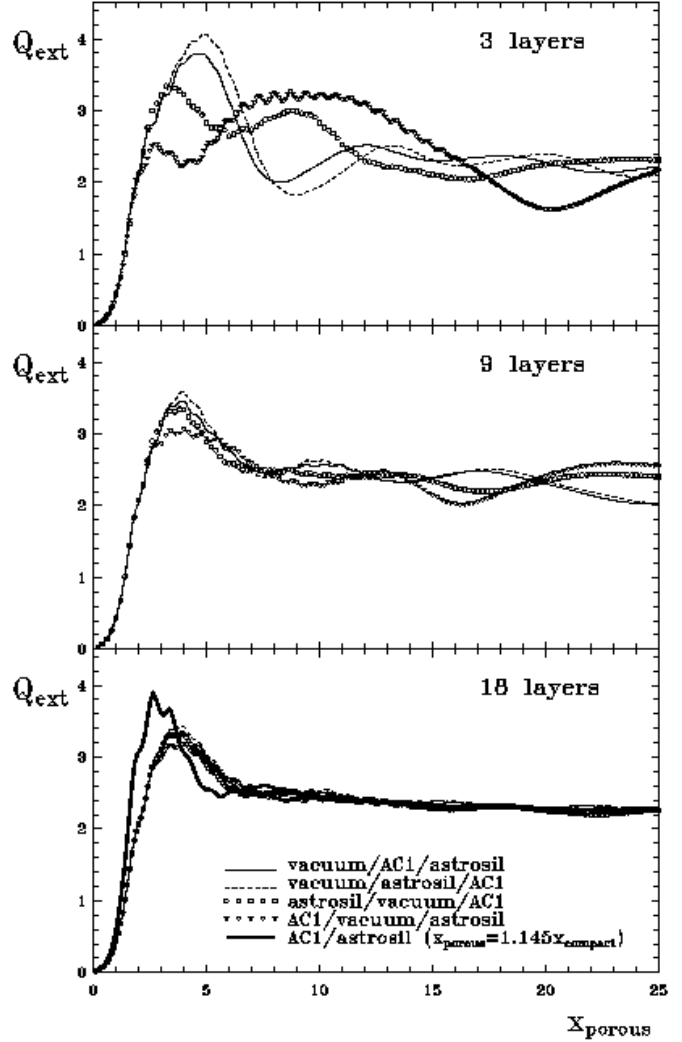}}
\caption{Size dependence of the extinction efficiency factors
for {\bf\rm layered} spherical particles.
Each particle contains an equal fraction
of amorphous carbon (AC1), astrosil and  vacuum {\bf\rm (the porosity ${\cal P}=1/3$)}
separated in equivolume layers.
The cyclic order of the {\bf\rm different material} layers is indicated (starting from the core).
The effect of the increase of the number of layers is illustrated.
The thick line at the lowest panel corresponds to compact spheres
consisting of AC1 and astrosil. For a given value of the size parameter,
the compact and porous particles have the same {\bf\rm mass.}
}\label{f1}
\ec
\end{figure}

We start with the consideration of layered particles.
Figure~\ref{f1} shows the extinction efficiency factors
$Q_{\rm ext}=C_{\rm ext}/ \pi r^2_{\rm s}$
($C_{\rm ext}$ is the extinction cross section,  $r_{\rm s}$
{\bf\rm the outer radius) of} layered spheres.
The optical properties of core-mantle spheres have been studied rather well
and {\bf\rm seem} to show no significant peculiarities\footnote{Except for
a resonance peak arising for particles with {\bf\rm mantles}
{\bf\rm having a} large refractive index (see, e.g., Gurwich et al.~\cite{gkso03}).}
(Babenko et al.~\cite{bak03}).
In contrast, three-layered spheres already
can produce  anomalous extinction of light.
The order of the materials strongly affects
the behaviour of extinction for such particles
(the upper panel of Fig.~\ref{f1}). First of all, the {\bf\rm location} of vacuum
(the core or the middle layer) is important.
The  curve {\bf\rm for} particles
with a carbon core and an outermost astrosil layer is the most peculiar curve.
Here, a very rare situation is observed:
the first maximum is damped,
but there is a very broad second {\bf\rm maximum.}
Note that the scattering efficiency depends more strong on
the order of layers than the absorption efficiency.
However, all the peculiarities
disappear {\bf\rm when} the number of layers increases: the difference
between the curves becomes rather small for particles with 9 layers (3 shells)
and is hardly present for particles with 18 layers
(6 shells; see  Figs.~\ref{f1} and \ref{f2a}).
Figure \ref{f2a} shows the size dependence of the scattering
($Q_{\rm sca}$)
and absorption
($Q_{\rm abs}$)
efficiency factors,
albedo $\Lambda=Q_{\rm sca}/Q_{\rm ext}$
       and the parameter $g^($\footnote{The notation $g$ is related to
         the Henyey--Greenstein phase function that is
         very often use in radiative transfer modelling.}$^)$
       describing the asymmetry of the phase function $F(\Theta, \Phi)$
       \be\label{g}
       g = \langle \cos \Theta \rangle = \frac
       {\int_{4 \pi} F(\Theta, \Phi)\, \cos \Theta \, {\rm d}\omega}
       {\int_{4 \pi} F(\Theta, \Phi)\, {\rm d}\omega}
       \ee
       for multi-layered spheres.
{\bf\rm As has been} noted by Voshchinnikov \&  Mathis~(\cite{vm99}):
{\bf\rm the optical properties weakly depend on the order of materials
in each of the shells and are close to some ``average'' properties,
if {\bf\rm a particle consists of} many shells ($\ga 3 - 5$).
In other words,}
for such particles, different efficiency factors as well as
albedo $\Lambda$ and
the asymmetry parameter $g$ 
depend {\bf\rm practically only} on the volume fraction of materials.

\begin{figure*}[!ht]\bc
\resizebox{\hsize}{!}{\includegraphics{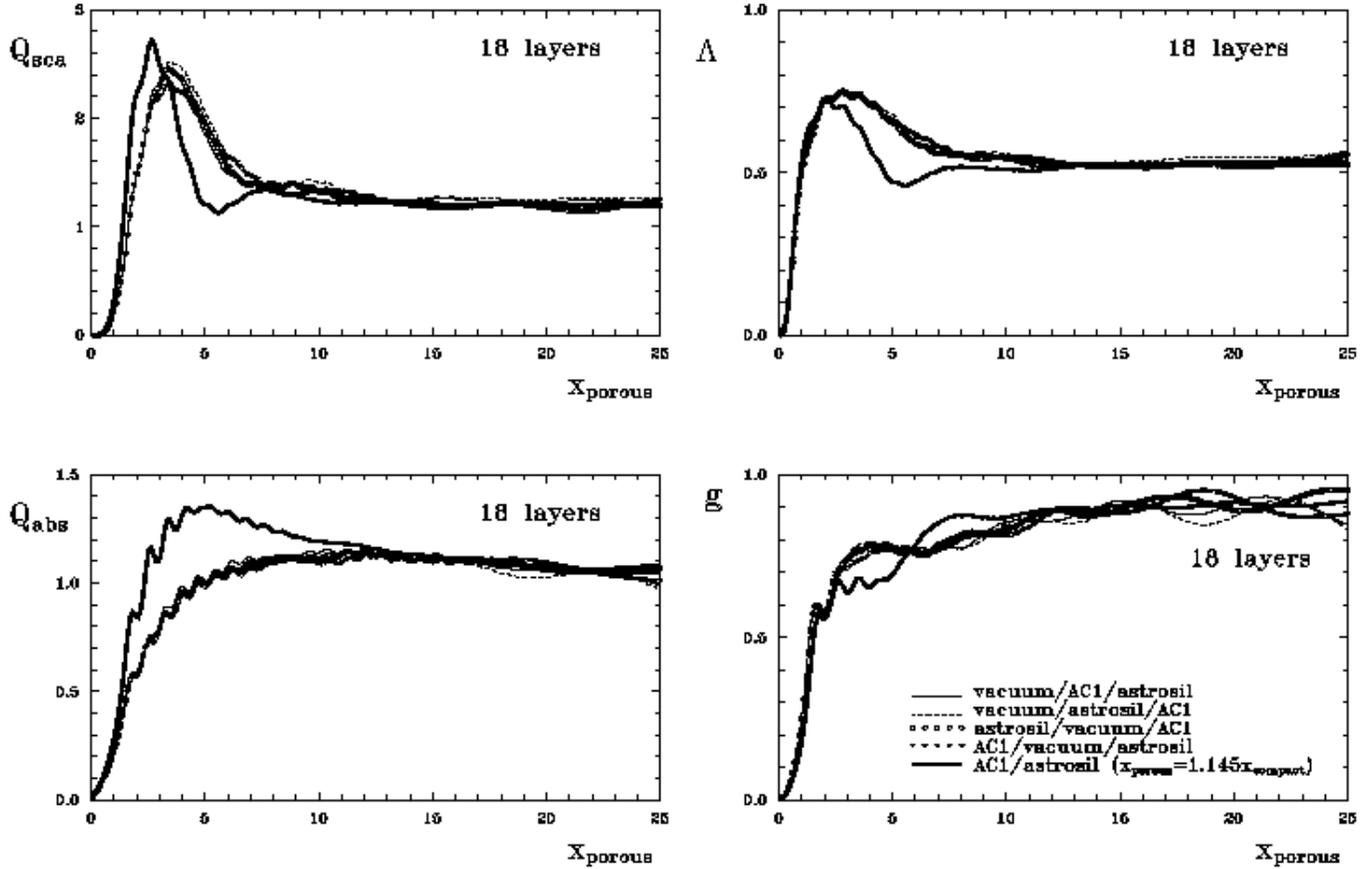}}
\caption
{Size dependence of the scattering and absorption efficiency factors,
albedo and asymmetry parameter for multi-layered spheres.
The parameters of particles are the same as in Fig.~\ref{f1}.}
\label{f2a} \ec \end{figure*}


{\bf\rm The} solid thick lines {\bf\rm in} Figs.~\ref{f1} and \ref{f2a}
show the size dependence of the optical characteristics {\bf\rm of} compact
(${\cal P} = 0$) spheres consisting of the same amount of solid materials
{\bf\rm as the porous ones}. To compare the
optical properties of porous and compact particles,
it is helpful to normalize the size parameter
{\bf\rm of either} the compact or porous particle using the relation
\be
x_{\rm porous} = \frac{x_{\rm compact}}{(1-{\cal P})^{1/3}}
= \frac{x_{\rm compact}}{(V_{\rm solid} /V_{\rm total})^{1/3}}. \label{xpor}
\ee
 In the case of the particles presented in Figs.~\ref{f1} and \ref{f2a},
this leads to stretching of the $x$ scale for compact particles
by a factor of $\sqrt[3]{3/2} \approx 1.145$.
 It can be seen that the presence of vacuum inside {\bf\rm the composite particles}
reduces the peak of the absorption efficiency (Fig.~\ref{f2a}, left lower panel)
and shifts  that of the scattering efficiency (Fig.~\ref{f2a}, left upper panel).
 Correspondingly,  these two effects explain
the behaviour of the curves for the extinction 
(Fig.~\ref{f1}).
 A medium porosity  influences the albedo and the
asymmetry parameter only in a restricted range of  size parameters.

Similar calculations for particles with inclusions {\bf\rm were performed many times using}
{\bf\rm the} DDA technique. But {\bf\rm so far} only Wolff
et al.~(\cite{wcms-l94}, \cite{wcg98}) considered  particles
with large (non-Rayleigh) inclusions. They computed the optical properties of
silicate spheres and spheroids with a size parameter up to
$x_{\rm porous} = 10$ and a volume fraction of vacuum inclusions up to 80\%.
For the efficiency factors and asymmetry parameter,
the {\bf\rm difference}
between particles with Rayleigh\footnote{\bf\rm In this case, the
size of the inclusion is much smaller than the radiation wavelength
in the medium.}
 and non-Rayleigh {\bf\rm vacuum} inclusions
{\bf\rm became} noticeable {\bf\rm for} the porosity ${\cal P} \ga 0.4$.

The results of our DDA calculations of the extinction efficiency factors
for pseudospheres with the porosity ${\cal P} =0.33$ are shown in
Figs.~\ref{f1i1}--\ref{f1i3}.
The {\bf\rm volume fractions of the materials are} approximately
the same as in {\bf\rm the particles presented in} Figs.~\ref{f1} and \ref{f2a} but the
materials are present in the form of inclusions instead of layers
(vacuum is considered as a matrix).
 Figure~\ref{f1i1} shows the results {\bf\rm obtained}
for particles with {\bf\rm single size inclusions}.
 {\bf\rm Note that despite} the
different structure of the targets (the  number
of inclusions {\bf\rm reduces} from $\sim 240000$ to $\sim 330$ {\bf\rm with growing $n_{\rm dip}$}),
the differences between the extinction efficiencies are quite small.
The same conclusion is correct for
particles {\bf\rm having} inclusions of different sizes.
 The results for three
different targets with a {\bf\rm distribution of inclusions of five sizes} are shown
{\bf\rm in Fig.~\ref{f1i2}}.
 {\bf\rm The size of the inclusions ranges from 1 to 9 and
the number of inclusions is inversely proportional to their volume.
 In other words, the total numbers of dipoles in the inclusions
of each size are approximately the same.}

The extinction efficiencies of the
particles with inclusions are compared
with those of {\bf\rm layered} particles {\bf\rm in} Fig.~\ref{f1i3}.
{\bf\rm The} difference between compact and porous particles
is clearly seen,
but the results for porous particles with inclusions
and layers {\bf\rm look} rather similar (excluding, perhaps, the height of {\bf\rm the} first
maximum).
This behaviour is confirmed by
Fig.~\ref{f2i} where {\bf\rm other efficiencies}, albedo and
the asymmetry parameter are presented.
 The largest deviations occur for the scattering {\bf\rm efficiency} $Q_{\rm sca}$
(in the range $x_{\rm porous} \approx 3 - 10$) and the asymmetry parameter
$g$ for {\bf\rm the size parameter $x_{\rm porous} \ga 10$}.
The latter does not seem to be an artifact related to small number of angles
used in our calculations during the averaging over scattering.

Thus, we can conclude that
if materials are ``well mixed'' inside a particle of intermediate porosity,
its internal structure in form of layers, Rayleigh or non-Rayleigh inclusions
{\it hardly} can be inferred from the transmitted radiation.
In contrast, there is a clear difference between the optical properties
of compact and porous grains.
\begin{figure}\bc
\resizebox{\hsize}{!}{\includegraphics{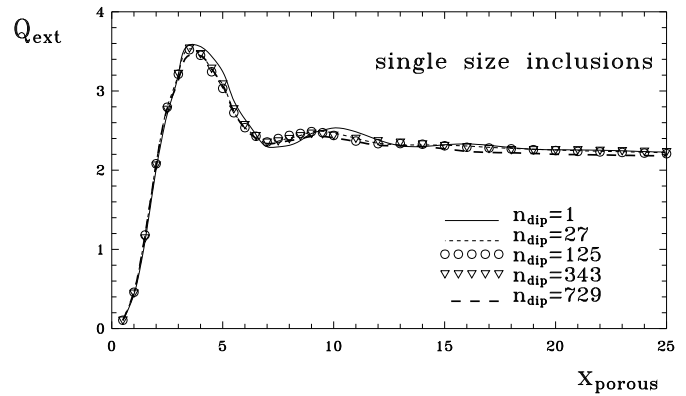}}
\caption{Size dependence of the extinction efficiency factors
for pseudospheres with inclusions
of the same single size {\bf\rm after averaging
of three different targets}.
Each particle contains an equal volume fraction (33.33\%)
of AC1, astrosil and  vacuum.
The effect  of variations of the size of inclusions is illustrated.
}\label{f1i1}
\ec
\end{figure}
\begin{figure}\bc
\resizebox{\hsize}{!}{\includegraphics{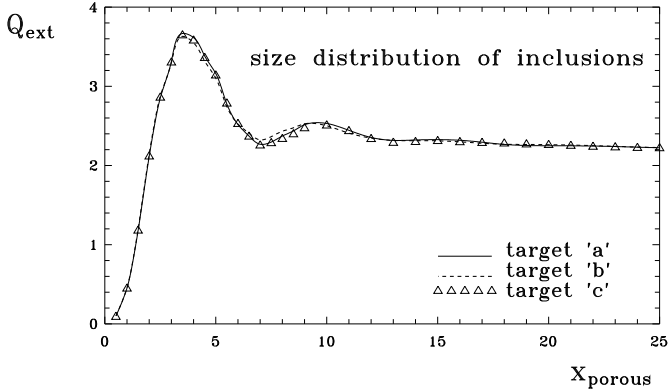}}
\caption{Size dependence of the extinction efficiency factors
for pseudospheres with  a size  distribution of inclusions.
Each particle contains an equal volume fraction (33.33\%)
of AC1, astrosil and  vacuum.
The volume fractions of inclusions of different sizes
are approximately the same.
}\label{f1i2}
\ec
\end{figure}

\begin{figure}\bc
\resizebox{\hsize}{!}{\includegraphics{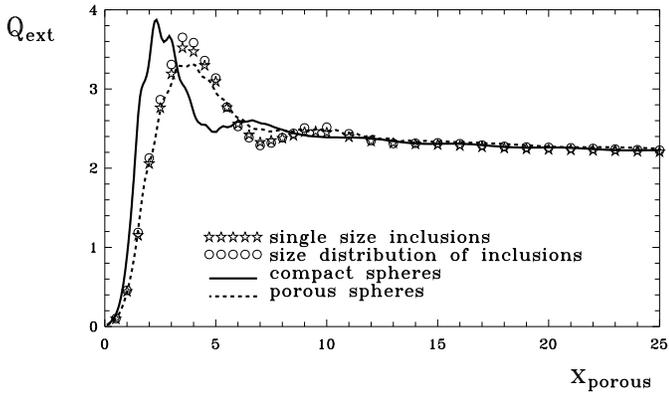}}
\caption{
The average extinction efficiencies
for particles with single size inclusions (from Fig.~\ref{f1i1})
and particles with size distribution of inclusions (from Fig.~\ref{f1i2}).
Each particle contains an equal volume fraction (33.33\%)
of AC1, astrosil and  vacuum.
The thick solid line corresponds to compact spheres
consisting of AC1 and astrosil. For a given value of the size parameter,
the compact and porous particles have the same mass.
 The thick dashed line shows the extinction for layered spheres
{\bf\rm after} averaging over four samples presented at the bottom panel of
Fig.~\ref{f1}.
}\label{f1i3}
\ec
\end{figure}


\begin{figure*}\bc
\resizebox{\hsize}{!}{\includegraphics{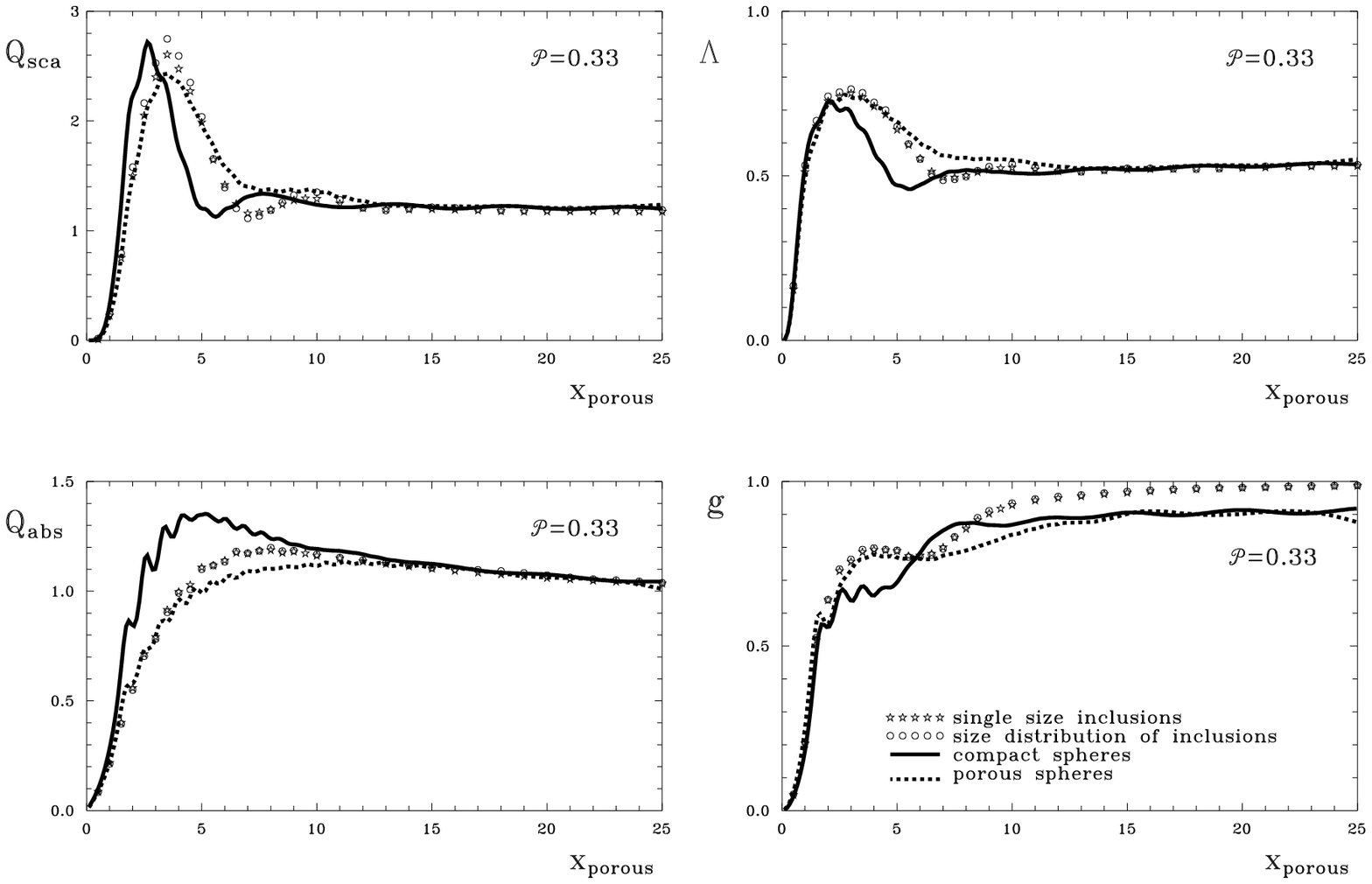}}
\caption
{Size dependence of the scattering and absorption efficiency factors,
albedo and {\bf\rm the} asymmetry parameter for pseudospheres with inclusions.
The parameters of particles are the same as in Fig.~\ref{f1i3}.}
\label{f2i} \ec \end{figure*}

\subsection{Very porous particles}\label{vac}

It is commonly accepted that the fraction of vacuum in interstellar
dust grains can be large.
{\bf\rm For example}, very porous particles are often used to model cometary grains
and dust in protoplanetary discs.
Greenberg \& Hage~(\cite{grha91}) {\bf\rm claim}
that the porosity of dust aggregates in comets can be in the range
$0.93 < {\cal P} < 0.98$. Their {\bf\rm conclusion is} based on the
model of porous grains developed by Hage \& Greenberg~(\cite{hagr90})
who used {\bf\rm a volume integral equation method similar to the DDA
for} light scattering calculations.
A verification of this method
had been made only for small compact spheres, but
the method was {\bf\rm applied to} large and very porous particles.
A qualitative agreement
between the results obtained with this method and from
Mie--Garnett calculations was found.
In both cases the
absorption {\bf\rm cross section} $C_{\rm abs}$ {\bf\rm increased and}
albedo $\Lambda$ {\bf\rm decreased} when the porosity {\bf\rm grew}.
 Although the validity of these conclusions
for particles beyond the Rayleigh domain remains unclear,
the results of Hage \& Greenberg~(\cite{hagr90}) are  frequently used for
estimates of grain properties in comets
(see, e.g., Mason et al.~\cite{masetal01}).
\begin{figure}\bc
\resizebox{\hsize}{!}{\includegraphics{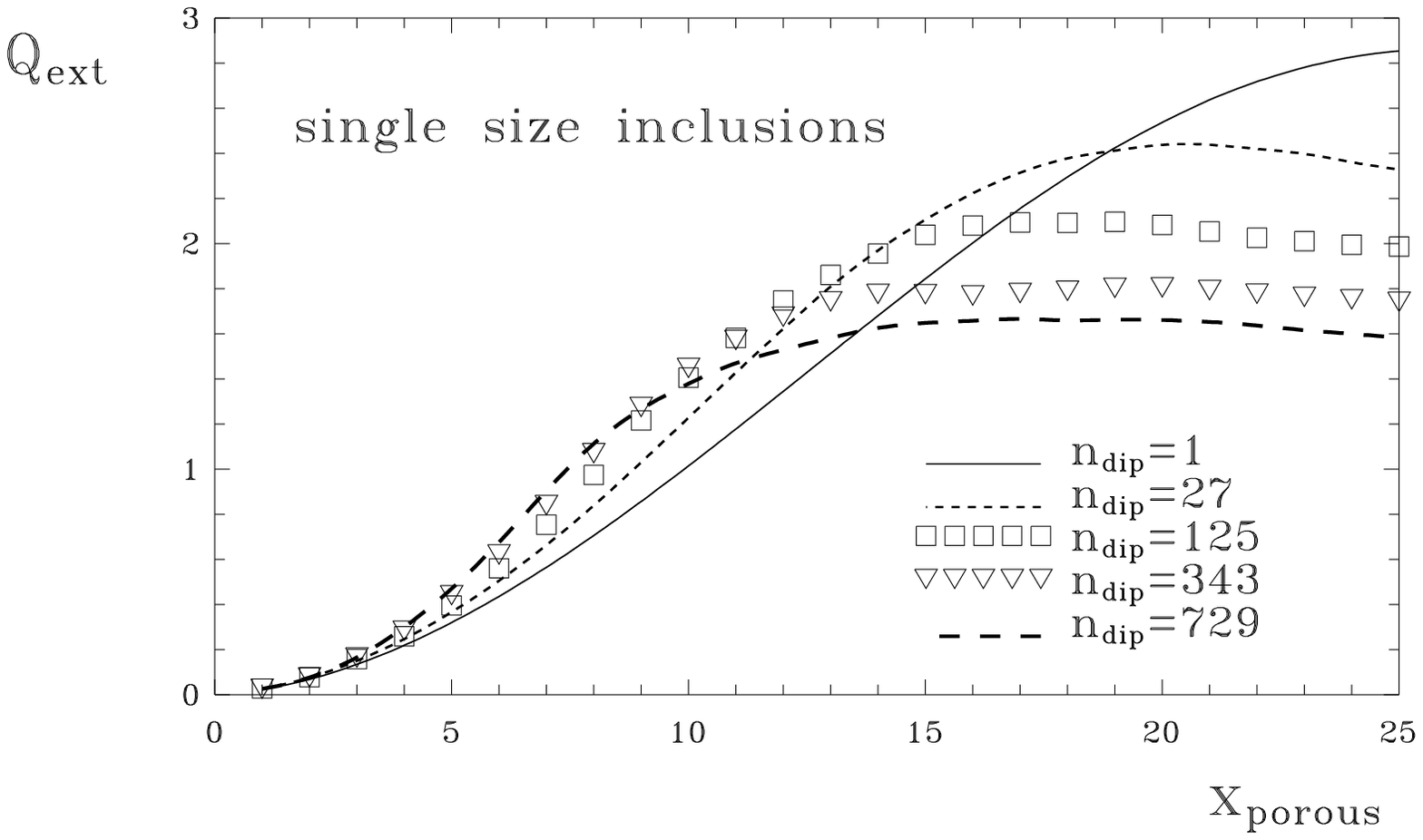}}
\caption{Size dependence of the extinction efficiency factors
for very porous pseudospheres with inclusions of {\bf\rm different} size.
Each particle contains  volume fractions
of AC1 and astrosil {\bf\rm equal to about 5\%, the {\bf\rm porosity} ${\cal P}=0.9$.}
The particles are similar to those presented in Fig.~\ref{f1i1}
but have larger porosity.
}\label{f09i1}
\ec
\end{figure}
\begin{figure}\bc
\resizebox{\hsize}{!}{\includegraphics{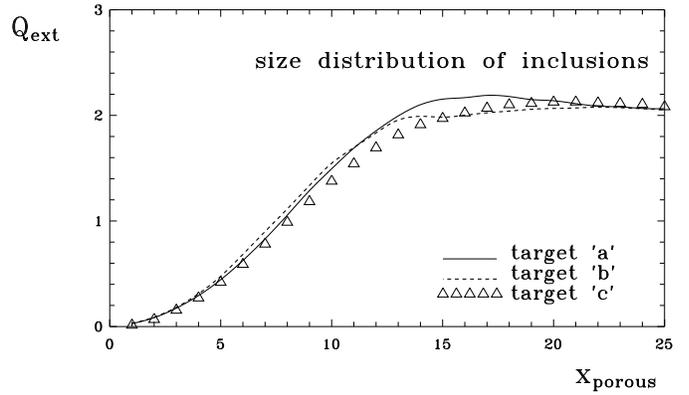}}
\caption{Size dependence of the extinction efficiency factors
for very porous pseudospheres with {\bf\rm a} size distribution of inclusions.
Each particle contains volume fractions
of AC1 and astrosil {\bf\rm equal to about 5\%, the {\bf\rm porosity} ${\cal P}=0.9$.}
The particles are similar to those presented in Fig.~\ref{f1i2}
but have larger porosity.
}\label{f09i2}
\ec
\end{figure}

\begin{figure}[ht]\bc
\resizebox{\hsize}{!}{\includegraphics{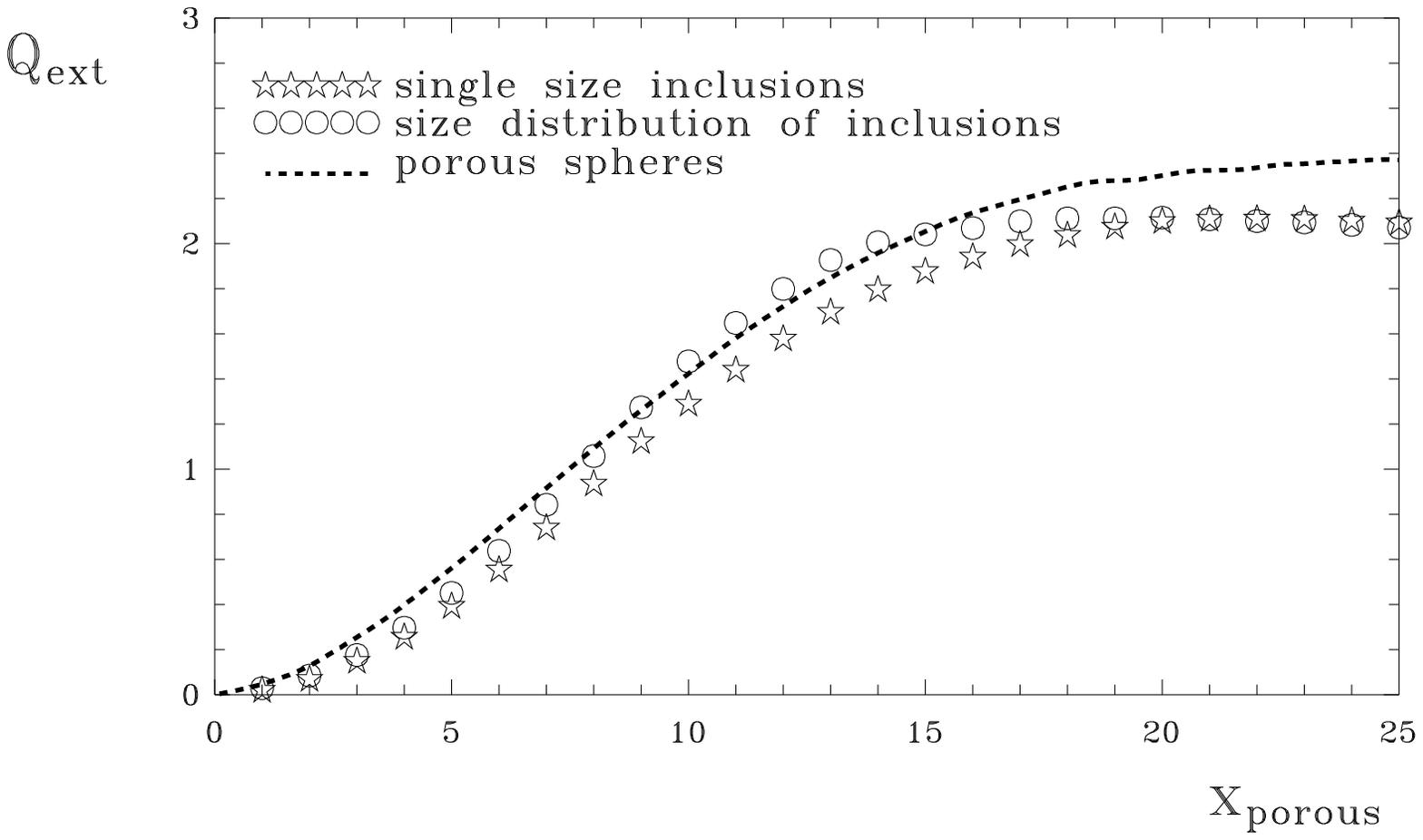}}
\caption{
Averaged extinction efficiencies
for particles with single size inclusions (from Fig.~\ref{f09i1})
and particles with {\bf\rm a} size distribution of inclusions (from Fig.~\ref{f09i2}).
Each particle contains  volume fractions
of AC1 and astrosil {\bf\rm equal to about 5\%, the {\bf\rm porosity} ${\cal P}=0.9$.}
{\bf\rm The particles are similar to those presented in Fig.~\ref{f1i3}
but have larger porosity.}
The thick dashed line shows the extinction for layered spheres.
}\label{f09i3}
\ec
\end{figure}


{\bf\rm Here} we {\bf\rm analyze} in detail {\bf\rm highly porous
particles in the case of ${\cal P}=0.9$}.
The results are presented in Figs.~\ref{f09i1}--\ref{f09i3} in {\bf\rm a} manner
similar to that {\bf\rm used} in Figs.~\ref{f1i1}--\ref{f1i3}.
{\bf\rm We should note the problem connected with the construction} of
targets {\bf\rm when} the size of inclusions {\bf\rm was} large.
In this case, the total number of dipoles from AC1 and astrosil {\bf\rm was}
$\sim 36000 - 39000$ and they {\bf\rm were} located in $\sim 50$ inclusions
which {\bf\rm might not always touch others}.
 Nevertheless, the changes of the general behaviour of extinction with
an increase of the number of dipoles in inclusions are clearly seen in
Fig.~\ref{f09i1}.
The extinction efficiency factors $Q_{\rm ext}$  become larger for small values of
$x$ and smaller for large ones.
  This essentially deviates from what was observed
for particles of intermediate porosity {\bf\rm (cf. Fig.~\ref{f1i1})}.
 The growth of porosity leads to the disappearance of the first maximum.
But the curves for particles with inclusions of large sizes
do not approach  the limiting value $Q_{\rm ext}=2$ defined
by the ``extinction paradox''.
This fact should be related to  special topology of very porous
particles with large inclusions (values of $n_{\rm dip}$).
Possibly, some inclusions intercept a part of  light
 scattered
from other inclusions and scatter it in the forward direction.
This  decreases the extinction.

\begin{figure*}\bc
\resizebox{\hsize}{!}{\includegraphics{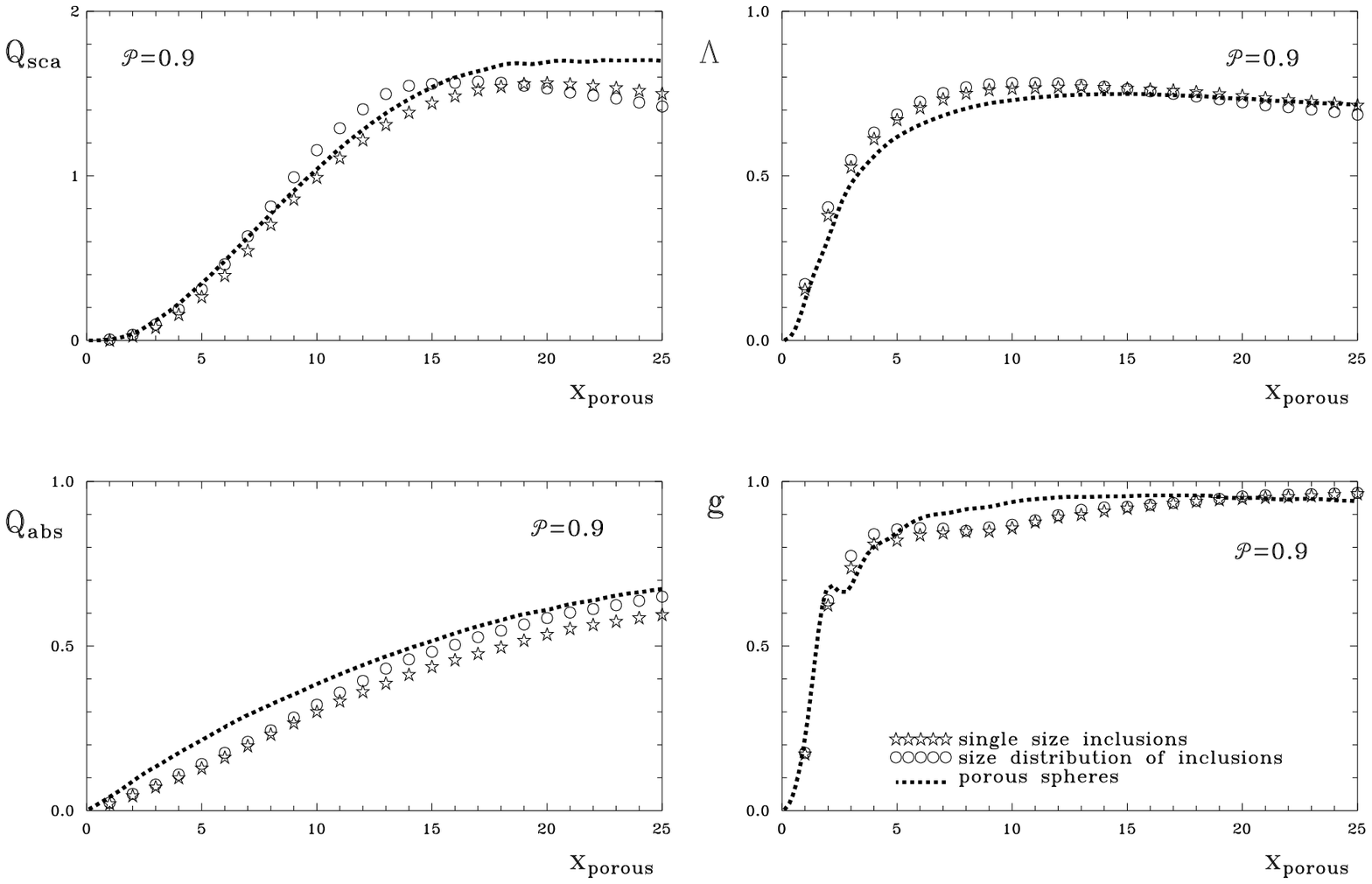}}
\caption
{Size dependence of the scattering and absorption efficiency factors,
albedo and {\bf\rm the} asymmetry parameter for pseudospheres with inclusions.
The parameters of particles are the same as {\bf\rm in} Fig.~\ref{f09i3}.}
\label{f209i} \ec \end{figure*}

However, if particles are composed of subparticles of different size,
their extinction is similar to the
usual one and the factors $Q_{\rm ext} \rightarrow 2$ if
$x\rightarrow \infty$ (see Fig.~\ref{f09i2}).
The same conclusion is valid for
extinction produced by an ensemble of particles with
inclusions of single sizes presented in Fig.~\ref{f09i3}.
This Figure together with Fig.~\ref{f209i} {\bf\rm demonstrates a} very important result:
the optical properties of very porous layered
particles and particles with inclusions are  similar.
 Note that in both cases the models of particles were constructed in such a
manner that the materials inside them  were ``well mixed'', i.e.
the location of inclusions in the form of layers or
islands is not distinguished.

 This leads to the interesting conclusion that
a very simple computational model of multi-layered particles
seems to be of possible use in treating the optics of
composite grains.

\subsection{Particles of different porosity}\label{dif}

The role of porosity in dust optics can be properly analyzed using
the normalized cross sections
\bea
C^{\rm (n)} = \frac{C({\rm porous \, grain})}
{C({\rm  compact \, grain \, of \, same \, mass})} = \nonumber \\
\,\,\,\,\,\,\, (1-{\cal P})^{-2/3}\,  \frac{Q({\rm porous \, grain})}
{Q({\rm  compact \, grain \, of \, same \, mass})}. \label{cn}
\eea
The quantity $C^{\rm (n)}$ shows how porosity increases or decreases
the cross section.
Such an investigation has been performed by
Kr\"ugel \& Siebenmorgen~(\cite{ks94})
for absorption cross sections of small particles with $x < 1$.
They calculated
the effective optical constants of porous particles
using, in particular, the Bruggeman mixing rule, and
applied the Mie theory  to get  $Q_{\rm abs}$.
Kr\"ugel \& Siebenmorgen find that the cross sections
$C^{\rm (n)}_{\rm abs}$ increase {\bf\rm with ${\cal P}$} until ${\cal P} \la 0.6$
and then decrease.

\begin{figure}\bc
\resizebox{\hsize}{!}{\includegraphics{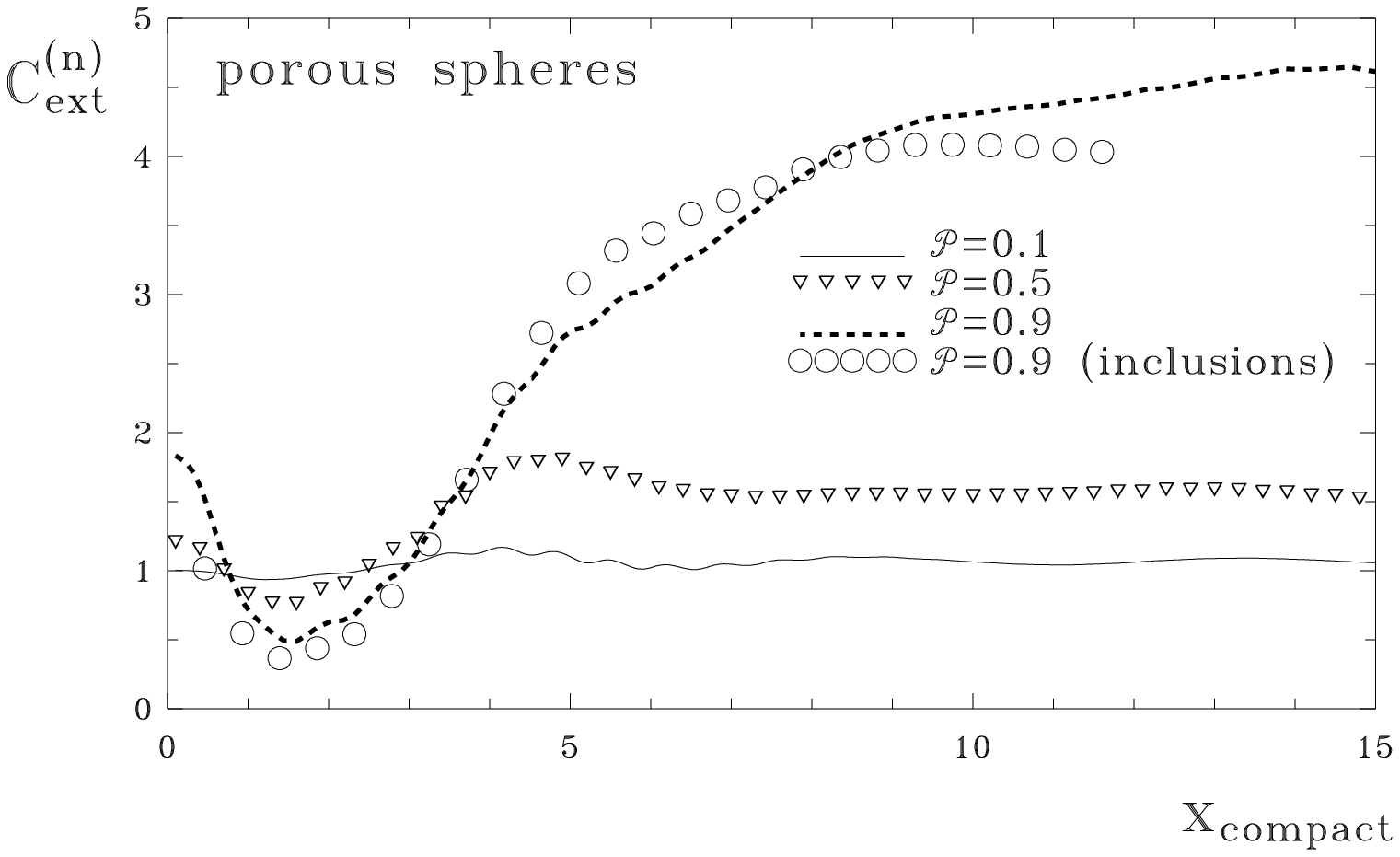}}
\caption
{
The  normalized extinction cross sections
for multi-layered spheres of different porosity.
Open circles {\bf\rm show} the normalized extinction for pseudospheres
with {\bf\rm a} size distribution of inclusions (see Fig.~\ref{f09i2}).
}
\label{fvac}\ec \end{figure}

Figure~\ref{fvac} shows the normalized extinction
cross sections  computed
for spheres of different porosity.
The results are plotted in the scale related to the size parameter
{\bf\rm $x_{\rm compact}$} calculated according to Eq.~(\ref{xpor}).
Since the extinction factors generally decrease when ${\cal P}$
increases (cf. the lower panel of Fig.~\ref{f1} and Fig.~\ref{f09i3}),
the values of $C^{\rm (n)}_{\rm ext}$
are greater than unity if the size parameter is smaller than $\sim 1$ or
larger than $\sim 3$. Thus, the  porosity increases the extinction of
small and large\footnote{For very large particles,
the normalized cross sections approach to asymptotic values
$C^{\rm (n)}  \rightarrow (1-{\cal P})^{-2/3}$
(see Eq.~(\ref{cn})) which are equal to 4.64 and 1.59 if
${\cal P} =0.9$ and 0.5, respectively.}
particles. An opposite case is observed only in a restricted range
of the size parameters $x_{\rm compact} \approx 1 - 3$
where the extinction by compact spheres has a maximum
(see, e.g., Fig.~\ref{f1i3}).

As follows from Fig.~\ref{fvac2}, such a behaviour
of the normalized extinction cross sections is  accompanied by
similar changes of the scattering and  absorption cross sections.
At the same time, the scattering and  absorption efficiencies
sharply and slightly grow with $x$ for very porous grains of large sizes.
 Note also that both $\Lambda$ (beginning
{\bf\rm with} $x_{\rm compact} \ga 2 - 3$)
and $g$ (for particles of all sizes) increase with {\bf\rm porosity}.
 The  behaviour of $C^{\rm (n)}_{\rm abs}$ and $\Lambda$
found by us is more complicated than predicted by Hage \& Greenberg~(\cite{hagr90}).
 Namely, the growth of porosity leads to an increase of $C^{\rm (n)}$
and {\bf\rm a} decrease of  $\Lambda$ for very small  size parameters,
and to an increase of both quantities for large values of $x$.
There exists also a small interval of intermediate size parameters where
both $C^{\rm (n)}$ and $\Lambda$ decrease.

Therefore, we can expect larger extinction, scattering and absorption
of radiation by porous particles with radius $r_{\rm s, compact}$
at wavelengths $\lambda \la 2/3 \pi r_{\rm s, compact}$  and
$\lambda \ga 2 \pi r_{\rm s, compact}$  in  comparison with compact
particles of the same mass. At the intermediate wavelengths,
the compact particles absorb and scatter more radiation.
For example, the ``importance'' of compact grains in the production
of extinction is larger at the near-UV/visual range of wavelengths
($0.21 \mu{\rm m} \la \lambda \la 0.63 \mu{\rm m}$)
and at the near-IR wavelengths
($2.1 \mu{\rm m} \la \lambda \la 6.3 \mu{\rm m}$) if
$r_{\rm s, compact}=0.1 \mu{\rm m}$ and
$r_{\rm s, compact}=1 \mu{\rm m}$, respectively.
Note that the latter estimates are rather approximate because of
the wavelength dependence of the refractive index of materials
(see Voshchinnikov et al. \cite{vihd03} for more details).

\begin{figure*}\bc
\resizebox{\hsize}{!}{\includegraphics{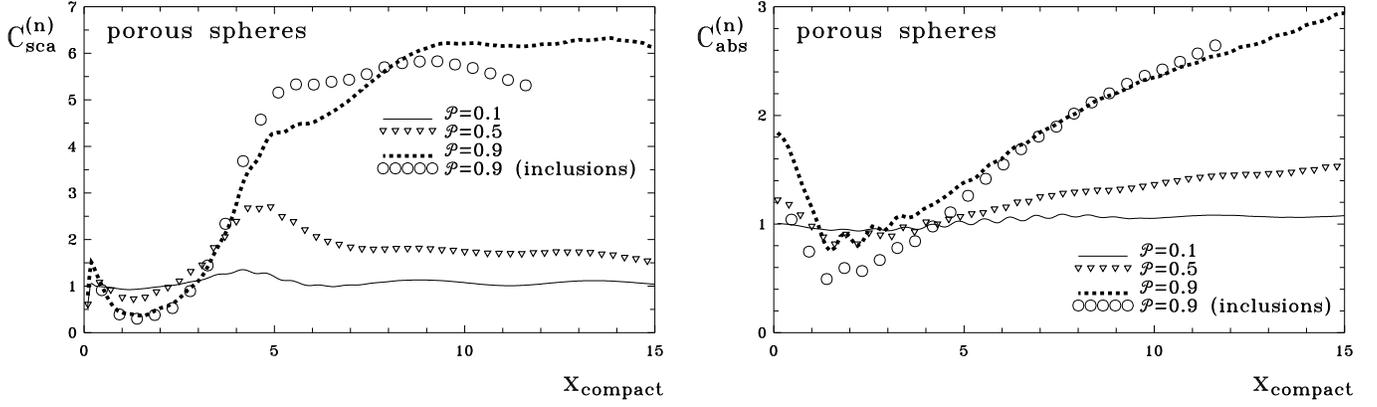}}
\caption
{Size dependence of the scattering and absorption normalized cross sections
for multi-layered porous spheres
and pseudospheres with {\bf\rm a} size distribution of inclusions.
The parameters of particles are the same as in Fig.~\ref{fvac}.
}\label{fvac2}\ec \end{figure*}

\section{Comparison with Effective Medium Theory}\label{emt}

\begin{figure*}[htb]\bc
\resizebox{\hsize}{!}{\includegraphics{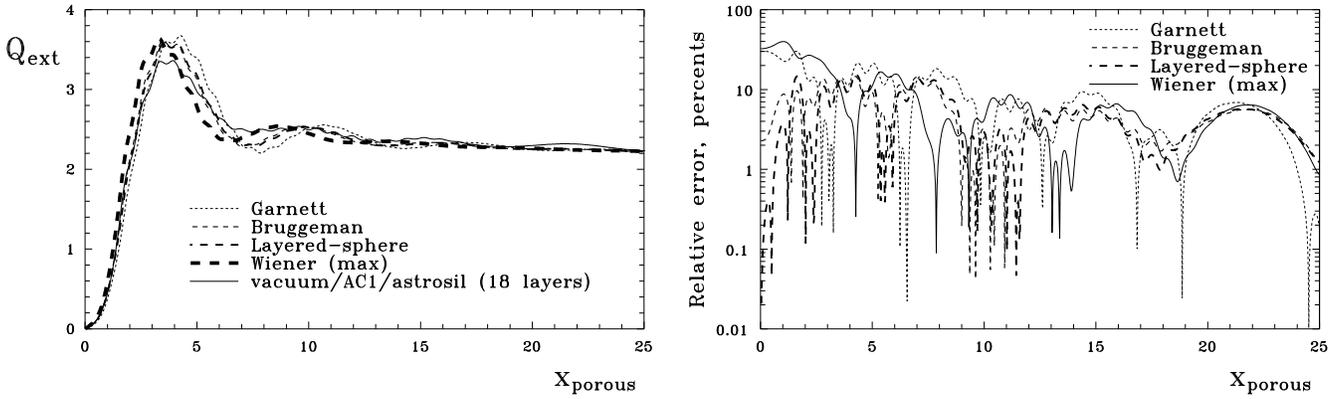}}
\caption{
   Size dependence of the efficiency factors (left panel)
   and their relative errors (right panel)
   calculated with the exact theory for multi-layered spheres
   and with the Mie theory using four different EMT rules.
   Multi-layered particles contain an equal {\bf\rm volume} fraction  of
   amorphous carbon (AC1), astrosil and vacuum.
   The cyclic order of the 18 layers is indicated.
}\label{emt1}\ec\end{figure*}

The EMT is an approach to treat inhomogeneous scatterers as
homogeneous particles having an average (effective)  refractive index.
The attempts to find the ``best'' EMT mixing rule {\bf\rm continue} up to
now (e.g., Maron \& Maron~\cite{mm04}).
The EMT is well described in the recent reviews of
Sihvola~(\cite{sih99}), Ch\'ylek et al.~(\cite{cval00})
and  papers of Spanier \& Herman~(\cite{sh00}) and
Kolokolova \& Gus\-taf\-son~(\cite{kolgust01}).
 There are many EMT rules, but besides a few
ones they are rather similar in principle.
 Here we give  formulas of the most often used EMT rules
for $n$-component mixtures:
the Garnett~(\cite{gar04}) and Bruggeman~(\cite{brug35}) rules.
In the first case, the mixing rule averages the
dielectric permittivities of inclusion materials $\ve_{i}$
and a ``matrix'' (host) material
$\ve_{\rm m}$$^($\footnote{The dielectric permittivity
is related to the refractive index as $\ve=m^2$.}$^)$
\be
{\ve}_{\rm eff} = \ve_{\rm m} \left( 1 +
\frac{ 3 \sum_{i} f_{i}
\displaystyle\frac{\ve_{i} - \ve_{\rm m}}{\ve_{i} + 2 \ve_{\rm m}}}
{ 1 - \sum_{i} f_{i}
\displaystyle\frac{\ve_{i} - \ve_{\rm m}}{\ve_{i} + 2 \ve_{\rm m}}}
\right),
\ee
where $f_{i}=V_i /V_{\rm total}$ is the volume fraction
of the $i$th material and $\ve_{\rm eff}$ is the effective permittivity.
The expression for the Bruggeman~(\cite{brug35}) rule is
\be
\sum_i f_{i} \frac{\ve_{i} - \ve_{\rm eff}}{\ve_{i} + 2 \ve_{\rm eff}} = 0.
\ee
As an example of a more sophisticated rule,
we use the ``layered-sphere EMT'' introduced by
Voshchinnikov \&  Mathis~(\cite{vm99}).
In this case, the effective optical constants $\ve_{\rm eff}$
are defined as follows (see also Farafonov \cite{f00}):
\be
\ve_{\rm eff} = \frac{1+2\alpha/V}{1-\alpha/V} =
   \frac{{\cal A}_2}{{\cal A}_1}, \label{ls1}
\ee
where $\alpha$ is the complex electric polarizability and
the coefficients ${\cal A}_1$ and ${\cal A}_2$ are obtained
as a result of multiplication of matrices depending on the optical
constants and volume fractions {\bf\rm of layers}
\bea
\left( \begin{array}{c}
{\cal A}_1 \\ {\cal A}_2
\end{array} \right) & = &
\left( \begin{array}{cc}
1 & {1}/{3} \\
\ve_n & -{2}/({3}\ve_n)
\end{array} \right)  \nonumber \\
& \times & \prod_{i=2}^{n-1}
\left( \begin{array}{cc}
{1}/{3} \left( \frac{\ve_{i}}{\ve_{i+1}} + 2 \right) &
-{2}/({9 f_{i}}) \left( \frac{\ve_{i}}{\ve_{i+1}} - 1 \right) \\
-f_{i} \left( \frac{\ve_{i}}{\ve_{i+1}} - 1 \right) &
{1}/{3} \left( 2 \frac{\ve_{i}}{\ve_{i+1}} + 1 \right)
\end{array} \right)
\nonumber \\
& \times &
\left( \begin{array}{c}
{1}/{3} \left( \frac{\ve_{1}}{\ve_{2}} + 2 \right) \\
- f_{i} \left( \frac{\ve_{1}}{\ve_{2}} - 1 \right),
\end{array} \right). \label{ls2}
\eea

The absolute bounds to $\ve_{\rm eff}$ were given by
Wiener (\cite{wiener10})\footnote{\bf\rm\rm These expressions were exactly derived
for non-absorbing materials but it seems they can be applied to slightly
absorbing materials too.}
\be
\ve_{\rm eff, \, max} = \sum_i f_i \ve_i,
\label{w1}
\ee
and
\be
{\ve_{\rm eff, \, min}} = \left( \sum_i \frac{f_i}{\ve_i} \right)^{-1}.
\label{w2}
\ee
For any {\bf\rm dielectric particle} composition and structure,
$\ve_{\rm eff}$ cannot lie beyond these limits
as long as the microstructure dimensions remain small compared with
the radiation wavelength.
Note that in the composite grain model of Mathis \&  Whiffen~(\cite{mw89})
the effective refractive index was calculated from Eq.~(\ref{w1}),
i.e. {\bf\rm a} maximum  of {\bf\rm possible} refractive indices was taken.

The general condition of EMT applicability is
that the size of ``inclusions'' (in the EMT the particle inhomogeneity
is considered in the form of uniformly distributed {\bf\rm small} inclusions)
is small in comparison to
the wavelength of incident radiation (Ch\'ylek et al. \cite{cval00}).
The real range of applicability of different rules was shown
to be nearly the same (see, e.g., Table~4 in Voshchinnikov \cite{v02}).

\begin{figure}[htb]\bc
\resizebox{\hsize}{!}{\includegraphics{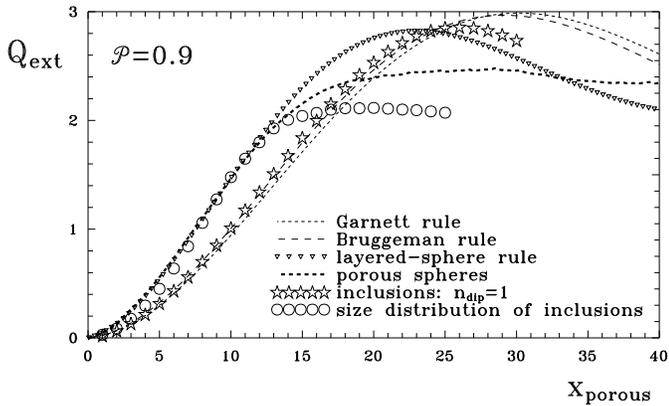}}
\caption{
   Size dependence of the extinction efficiency factors
   calculated for multi-layered spheres, pseudospheres with inclusions
   and with the Mie theory using three different EMT rules.
   The porosity of particles is ${\cal P}=0.9$.
}\label{emt2}\ec\end{figure}

Now let us discuss how different EMT rules can reproduce
the optical properties of layered spheres.
Figure~\ref{emt1} (left panel)  shows the extinction efficiency factors computed
with the exact theory for layered spheres
and with the Mie theory  using
Garnett\footnote{Vacuum was adopted as the matrix material.},
Bruggeman, and layered-sphere mixing rules of
the EMT {\bf\rm (i.e. with the effective refractive indices are equal to
$m_{\rm eff}=1.496+0.060i$,
$m_{\rm eff}=1.541+0.081i$ and
$m_{\rm eff}=1.529+0.080i$, respectively)}.
Wiener's  maximum bound is $m_{\rm eff}=1.604+0.105i$.
Figure~\ref{emt1} (right panel) demonstrates the relative errors
of these EMTs.
It can be seen that the errors of the Bruggeman and layered-sphere rules
are  of several percent or better in the considered range of particle sizes.
The same is generally true for other
efficiency factors and albedo.
 An exception is the region after the first maximum of
the scattering efficiency factor and albedo ($x_{\rm porous} \approx 6 - 8$)
where the relative errors may reach up to 20\%. The largest errors
occur for the asymmetry parameter, especially for small size parameters.
 The high accuracy of the layered-sphere rule in the case of
 very small particles sizes
is explained by the fact that it is based on the Rayleigh approximation.

 Other rules of the EMT behave like the Garnett and Bruggeman
rules.
The general condition of
the EMT applicability is not fulfilled for layered particles
as ``inclusions'' (layers) are not small in comparison with
the wavelength of incident radiation.
However, most rules of the EMT
can reproduce the optical properties of layered
spheres of any size, if the number of layers is larger than 15~--~20
(and the particles with well mixed inclusions
as it was shown in Sect.~\ref{mpor}).
This conclusion, however, can be affected by the porosity of particles.

 Figure~\ref{emt2}  illustrates the applicability of different EMT rules
to particles of very high porosity.
The cases of other efficiency factors, albedo and asymmetry parameter
are similar. The Figure demonstrates that the layered-sphere rule
rather well reproduces the optical properties of layered spheres as well as
the particles with inclusions of different sizes
(the errors  are smaller than 10~--~20\% if
$x_{\rm porous} \la 15$ and ${\cal P}=0.9$).
Note that for such particles other  rules provide acceptable
approximations for intermediate porosity (${\cal P} \la 0.5$).
The  Garnett and Bruggeman rules together with
the Mie theory  rather well approximate the light scattering by
particles with inclusions of small sizes.
Therefore, all previous models based on the standard EMT-Mie calculations
are related to particles composed of subparticles of very small sizes.
If the size of subparticles is not small,
only the layered-sphere rule {\bf\rm can} be used for {\bf\rm the} description
of the optical properties of {\bf\rm very} porous scatterers.

\section{Concluding remarks}\label{concl}

We consider {\bf\rm different (including new)}
computational {\bf\rm approaches} to calculating the optical properties
of composite and porous grains
{\bf\rm that} can be used  for the interpretation of observations of  interstellar,
circumstellar and cometary dust.
 In our models the particles are represented by multi-layered
spheres or pseudospheres with inclusions of one or different sizes.
 If the number of layers is small,
 {\bf\rm  our model of layered spheres} coincides
with older models {\bf\rm of the grains having} several coatings.
 For a large ($\ga 15-20$) number {\bf\rm of} layers,
the model of layered spheres can approximate heterogeneous particles
consisting of  inclusions of different sizes.
 This gives us a handy way to treat composite grains {\bf\rm employing}
a very simple computational model of multi-layered particles instead of
time-consuming DDA calculations.

We make a careful examination of the optical
properties of particles of various porosity.
 Previously, this task was solved using the Mie theory
for homogeneous spheres and  effective refractive indices
derived from different mixing rules of the Effective Medium Theory.
 It is demonstrated that this approach gives {\bf\rm relatively accurate} results
{\bf\rm only if particles have} small (Rayleigh) inclusions.
 {\bf\rm Otherwise}, the approach {\bf\rm becomes} unacceptable
when the porosity exceeds {\bf\rm $\sim$0.5}.
 {\bf\rm An exception is provided by a sophisticated layered-sphere mixing rule,
recently suggested by Voshchinnikov \&  Mathis~(\cite{vm99}),
{\bf\rm that gives} results of acceptable accuracy
for particles with non-Rayleigh inclusions as well.
Note, however, that our consideration was restricted by spheres,
non very
absorbing materials and the integral scattering characteristics
but not the differential cross sections or elements of the
scattering matrix.

Some astrophysical applications of the model of layered grains
(in particular,
the possibility to reduce the model dust-phase abundances)
will be presented in a subsequent paper
(Voshchinnikov et al. \cite{vihd03}).
Further development of the model of  multi-layered particles will involve
consideration of non-spherical inhomogeneous grains
(see Farafonov et al. \cite{fip02}
for a review).

\acknowledgements{
We are grateful to Bruce Draine and Piotr Flatau for providing
DDSCAT 6.0 and to the referee for very careful reading of the paper
and recommendations related to its improvement.
The work was partly supported by  grants
of the DFG Research Group ``Laboratory Astrophysics''
and by grant 1088.2003.2 of the President of the Russian Federation
for leading scientific schools.
{\bf\rm V.I. acknowledges a support by the grant E02-11.0-8 of
the Russian Ministry of Education.}
}


\end{document}